\documentstyle{elsart}

\newlength{\defitemindent}
     {\begin{list}{}
               {\labelsep=0ex
            \settowidth{\labelwidth}{\hspace{\defitemindent} #1 \hspace{0.5ex}}
            \leftmargin=\labelwidth
            \addtolength{\leftmargin}{\labelsep}
             }}
     {\end{list}}

           {\arraycolsep=0.14em
                      \begin{eqnarray}}
                      {\end{eqnarray}}

\newenvironment{Eqnarray*}%
           {\arraycolsep=0.14em
                      \begin{eqnarray*}}
                      {\end{eqnarray*}}

\newcommand{\YOU}{\nonscript\kern 0.2ex U \nonscript\kern-0.2ex}
\let\member=\in

\newcommand{\sat}{\models}

\newcommand{\union}{\cup}

\renewcommand{\phi}{\varphi}

\newcommand{\imp}{\Rightarrow}

\newcommand{\G}{{\cal G}}
\newcommand{\I}{{\cal I}}

\newcommand{\R}{{\cal R}}

\newcommand{\ie}{i.e.,~}

\newcommand{\ol}{\setlength{\itemsep}{0pt}\begin{enumerate}}
\newcommand{\eol}{\end{enumerate}\setlength{\itemsep}{-\parsep}}
\newcommand{\ul}{\setlength{\itemsep}{0pt}\begin{itemize}}
\newcommand{\dl}{\setlength{\itemsep}{0pt}\begin{description}}
\newcommand{\edl}{\end{description}\setlength{\itemsep}{-\parsep}}
\newcommand{\eul}{\end{itemize}\setlength{\itemsep}{-\parsep}}

\newcommand{\subG}{_G}

\newcommand{\cI}{{\cal I}}

\newcommand{\EG}{E_G}
\newcommand{\CG}{C_G}

\newcommand\eqdef{=_{\rm def}}

\newcommand{\commentout}[1]{}

\newcommand{\bi}{\begin{itemize}}
\newcommand{\ei}{\end{itemize}}
\newcommand{\be}{\begin{enumerate}}
\newcommand{\ee}{\end{enumerate}}

\def\rarrowr{\buildrel{\smash{\raise 0.5ex \hbox{$\scriptstyle r$}}} \over
           {\smash{\mathop{\hbox to 1.3em {\rightarrowfill}}}} }

\newcommand{\gammafair}%
{\gamma^{{\it bt}}_{\mbox{\scriptsize{{\it fair}}}}}
\newcommand{\gammafairk}%
{\gamma^{{\it bt}}_{\mbox{\scriptsize{{\it fair,k}}}}}

\newcommand{\snt}{\mbox{{\it sent\/}}}
\newcommand{\eps}{\varepsilon}

\newcommand{\Ce}{C^\eps}
\newcommand{\Ee}{E^\eps}
\newcommand{\Cd}{C^\diamond}
\newcommand{\Ed}{E^\diamond}

\newcommand{\evi}{{\tt ev}_\cI}

\newcommand{\Ifmp}{\mbox{$\I^{\kern 0.1ex \it fm'}$}}

{\end{EXERCISE}}
{\end{HARDEX}}
{\end{SUPERHARD}}
\newcommand{\oldindex}[1]{}

\newcommand{\gloscg}{$C_G$}
\newcommand{\gloscgd}{$\Cd\subG$}

\newcommand{\gloseg}{$E_G$}
\newcommand{\glosegd}{$\Ed\subG$}
\newcommand{\glosege}{$\Ee\subG$}

\newcommand{\glospsie}{$\psi_e$}

\begin{document}
\begin{frontmatter}
\title{Common knowledge revisited\thanksref{book}}
\thanks[book]{This material appeared in the 1996 Conference on
Theoretical Aspects of Rationality and Knowledge.
It is based on our recently published book
\protect\cite{FHMV}.}
\author{Ronald Fagin\thanksref{ronthanks}}
\thanks[ronthanks]{email:~{\tt fagin@almaden.ibm.com};
URL:~{\tt http://www.almaden.ibm.com/cs/people/fagin/}}
\address{IBM Almaden Research Center,
650 Harry Road,
San Jose, CA 95120--6099}
\author{Joseph Y.\ Halpern\thanksref{joethanks}}
\thanks[joethanks]{This research was done while this author was at the
IBM Almaden Research Center.
Work supported in part by the Air Force Office of
Scientific Research (AFSC), under Contract F49620-91-C-0080.
email:~{\tt halpern@cs.cornell.edu};
URL:~{\tt http://www.cs.cornell.edu/home/halpern}}
\address{Department of Computer Science,
Cornell University,
Ithaca, NY 14853}
\author{Yoram Moses\thanksref{yoramthanks}}
\thanks[yoramthanks]{Part of this
research was performed while this author was on
sabbatical at Oxford. His work is
supported in part by a Helen and Milton A.\ Kimmelman career
development chair.
email:~{\tt yoram@cs.weizmann.ac.il};
URL:~{\tt http://www.wisdom.weizmann.ac.il/$^\sim$yoram}}
\address{Dept.~of Applied Math.\ and Comp.~Sci.,
The Weizmann Institute of Science,
Rehovot 76100, Israel}
\author{Moshe Y.\ Vardi\thanksref{moshethanks}}
\thanks[moshethanks]{This research was done while this author was at the
IBM Almaden Research Center.
email:~{\tt vardi@cs.rice.edu};
URL:~{\tt http://www.cs.rice.edu/$^\sim$vardi} }
\address{Dept.~of Computer Science,
Rice University,
Houston, TX 77005-1892}
\begin{abstract}
We consider the common-knowledge paradox raised
by Halpern and Moses:
common knowledge is necessary for
agreement and
coordination, but common knowledge
is unattainable in the real world because of temporal imprecision.
We discuss two solutions to this paradox:
(1) modeling the world with a
coarser granularity, and (2) relaxing the requirements for
coordination.
\end{abstract}
\end{frontmatter}

\section{Introduction}
The notion of {\em common knowledge\/},
where everyone knows, everyone knows that everyone knows, etc.,
has proven to be fundamental in
various disciplines,
including Philosophy \cite{Lew}, Artificial Intelligence \cite{MSHI},
Game Theory \cite{Au}, Psychology \cite{ClM}, and
Distributed Systems \cite{HM1}.
This key notion was first studied by the philosopher David
Lewis \cite{Lew}
in the context of conventions.  Lewis pointed out that in order
for something to be a convention, it must in fact be common knowledge
among the members of a group.  (For example, the convention that green
means ``go'' and red means ``stop'' is presumably common knowledge among
the drivers in our society.)

Common knowledge also arises in discourse understanding
\cite{ClM}.
Suppose Ann asks Bob ``What did you think of the movie?'' referring to
a showing of {\em Monkey Business\/} they have just seen.  Not only must
Ann and Bob both know that ``the movie'' refers to
{\em Monkey Business},
but Ann must know that Bob knows (so that she can
be sure that Bob will give a reasonable answer to her question),
Bob must know that
Ann knows that Bob knows (so that Bob knows that Ann will respond
appropriately to his answer), and so on.  In fact,
by a closer analysis of this situation, it can be shown that
there must be common knowledge of what movie is meant
in order
for
Bob to answer the question appropriately.

Finally,
as shown in \cite{HM1},
common knowledge also turns out to be
a prerequisite for agreement and
coordinated action.
This is precisely what makes
it such a crucial notion in the analysis of interacting groups of
agents.
On the other hand,
in
practical
settings common knowledge is impossible to achieve.
This puts us in a somewhat
paradoxical situation, in that we claim both that
common knowledge is
a prerequisite for agreement and coordinated action
and that it cannot be attained.
We discuss two answers to this paradox:  (1) modeling the world with a
coarser granularity, and (2) relaxing the requirements for
coordination.

\section{Two puzzles}
We start by discussing two well-known puzzles
that
involve
attaining common knowledge.  The first is the ``muddy children'' puzzle
(which goes back
at least to
\cite{Gam}, although the version we
consider here is taken from \cite{Bar}).

The story goes as follows:
Imagine $n$ children playing together.
Some, say $k$ of them,
get mud on their foreheads.  Each can see the mud on others but
not on his own forehead. Along comes
the father, who says,
``At least one of you has mud on your forehead,''
thus
expressing a fact known to each of them before he spoke (if $k>1$).  The
father then
asks the following question, over and over:
``Does any of you know whether you have mud on your
own forehead?''
Assuming that all the children are perceptive, intelligent,
truthful, and that they answer simultaneously, what will happen?

There is a straightforward proof by induction
that the first $k-1$ times he asks the question,
they will all say ``No,'' but then the $k^{\rm th}$ time the
children with muddy foreheads will all answer ``Yes.''
Let us denote the fact ``at least one child has a muddy forehead''
by~$p$.
Notice that if $k>1$, i.e., more than one child has a muddy forehead,
then every child can see at least one muddy forehead, and
the children initially all know $p$.
Thus, it would seem that the father does not provide the children
with any new information, and so
he should not need to tell
them that $p$ holds when $k>1$.
But this is false!
What the father provides is common knowledge.
If exactly $k$
children have muddy foreheads, then it is straightforward to see that
$E^{k-1} p$ holds before the father
speaks, but $E^k p$ does not
(here $E^k \phi$ means $\phi$, if $k=0$,
and everyone knows $E^{k-1} \phi$, if $k \ge 1$).
The father's statement actually converts the children's
state of knowledge from $E^{k-1} p$ to $C p$
(here $C p$ means that there is common knowledge of $p$).
With this extra knowledge, they can deduce whether their foreheads
are muddy.

In the muddy children puzzle, the children do not actually need common
knowledge;
$E^k p$ suffices for them to figure out whether they have mud
on their foreheads.
On the other hand, the {\em coordinated attack\/} problem
introduced by Gray
\cite{Gray}
provides an example where common
knowledge is truly necessary.
In this problem, two generals, each commanding a division of an army,
want to attack a common enemy.
They will win the battle only if they attack the enemy simultaneously;
if only one division attacks, it will be defeated.
Thus, the generals want to coordinate their attack.  Unfortunately, the
only way they have of communicating is by means of messengers,
who
might get lost or captured by the enemy.

Suppose a messenger sent by General~$A$ reaches General~$B$ with a
message saying ``{\sl attack at dawn}.''
Should General~$B$ attack?
Although the message was in fact
delivered, General~$A$ has no way of knowing that it was delivered.
$A$ must therefore consider it possible that $B$ did not receive the
message (in which case~$B$ would definitely not attack). Hence~$A$
will not attack given his current state of knowledge.
Knowing this, and not willing to
risk attacking alone, $B$ cannot attack based solely on
receiving~$A$'s message. Of course, $B$ can try to improve matters by
sending the messenger  back to~$A$ with an acknowledgment.
Even if the messenger reaches $A$, similar reasoning shows that neither
$A$ nor $B$ will attack at this point either.  In fact,
Yemini and Cohen \cite{YC} proved, by induction on the number of
messages, that
no number of
successful deliveries of acknowledgments to acknowledgments can
allow the generals to attack.
Halpern and Moses \cite{HM1} showed the relationship between
coordinated attack and common knowledge,
and used this to give a ``knowledge-based'' proof of Yemini and
Cohen's result.
Specifically,
assume that the generals behave according to some predetermined
deterministic protocol; that is, a general's actions (what messages he
sends and whether he attacks) are a deterministic function
of his history and the time on his clock.
Assume further that in the absence of any successful communication,
neither general will attack.
Halpern and Moses then prove the following theorem:
\begin{thm}{\rm \cite{HM1}}\label{attack}
A correct protocol for the coordinated attack problem must have the
property that whenever the generals attack, it is common knowledge
that they are attacking.
\end{thm}

Halpern and Moses then define the notion of a system where
{\em communication is not guaranteed}.
Roughly speaking,
this means
(1) it is always possible that from some point on, no messages will
be received, and (2) if a processor (or general)
$i$~does not
get any information to the contrary (by receiving some message), then
$i$~considers it possible that none of its messages were received.
In particular, in the coordinated attack problem as stated,
communication is not guaranteed.
Halpern and Moses
then prove that in such a system, nothing can become
common knowledge unless it is also common knowledge in the
absence of communication.
This implies the impossibility of coordinated attack:

\begin{thm}{\rm \cite{HM1}}\label{imposs}
Any correct protocol for the coordinated attack problem
guarantees that neither
general
ever attacks.
\end{thm}

Common knowledge of $\phi$ is defined to be the
infinite conjunction of the formulas $E^k\phi$.
This definition seems to suggest that common knowledge has
an ``inherently infinite'' nature.
Indeed, for a fact that is not common
knowledge to become common knowledge, each participating agent
must come to know an infinite collection of new facts.
Could this be one of the reasons that common knowledge
is impossible to attain in this case?
As we shall see,
it is not.

In practice, there is always a finite bound on the number
of possible local states of an agent in a real-world system.
A {\em finite-state system} is one where
each agent's set of possible local states is finite.
Fischer and Immerman \cite{FI1} showed that
in a finite-state system, common knowledge is
equivalent to~$E^k$ for a sufficiently large~$k$.
Nevertheless,
the result that common knowledge is not attainable if communication is
not guaranteed applies equally well to finite-state systems (as do
our later results on the unattainability of common knowledge).
Thus, in such cases, $E^k\phi$ is unattainable for some
sufficiently large~$k$.
(Intuitively, $k$~is large enough so that the agents cannot count up
to~$k$; that is, $k$~is
tantamount to infinity for these agents.)
So the unattainability of common knowledge in this case
is not due to the
fact that common knowledge
is defined in terms of an infinite conjunction.

\section{Common Knowledge and Uncertainty}
\label{ck-and-unc}
As we have seen,
common knowledge cannot be attained when
communication is not guaranteed.
Halpern and Moses show further that
common knowledge
cannot be attained
in a
system in which communication {\em is} guaranteed, but where
there is
no bound on the time it takes for messages to be delivered.
It would seem that when all messages are  guaranteed
to be delivered within a fixed amount of time,
say one
second,
attaining common knowledge should be a simple matter.
But things are not always as simple as they seem; even in
this case, uncertainty causes major difficulties.

Consider the following example:
Assume that two agents, Alice and Bob,
communicate over a channel in which (it is common knowledge that)
message delivery is guaranteed.
Moreover, suppose that there is only slight uncertainty concerning
message delivery times.
It is commonly known that any message sent from Alice to Bob
reaches Bob within $\eps$~time units.
Now suppose that at some point Alice sends Bob a message~$\mu$ that
does not specify the sending time in any way.
Bob does not know
initially that Alice sent him a message.
We assume that when Bob receives Alice's message, he knows that it
is from her.
How do Alice and Bob's state of knowledge
change with time?

Let $\snt(\mu)$
be the statement that Alice sent the message~$\mu$.
After $\eps$ time units, we have $K_A K_B \snt(\mu)$, that is,
Alice knows that Bob knows that she sent the message~$\mu$.
And clearly, this state of knowledge does not occur
before~$\eps$ time units.
Define
$(K_A K_B)^k \snt(\mu)$ by letting it be $\snt(\mu)$ for
$k=0$, and $K_A K_B (K_A K_B)^{k-1} \snt(\mu)$ for $k \ge 1$.
It is not hard to verify
that
$(K_A K_B)^k \snt(\mu)$ holds after
$k \eps$ time units, and does not hold before then.
In particular, common knowledge of $\snt(\mu)$ is never attained.
This may not seem too striking when we think
of~$\eps$ that is relatively large, say a day, or an hour.
The argument, however, is independent of the
magnitude of~$\eps$, and remains true
even for small values of~$\eps$.
Even
if
Alice and Bob are guaranteed that Alice's message
arrives
within one nanosecond, they still never
attain common knowledge that her message was sent!

Now let us consider what happens if
both Alice and Bob use the {\em same\/} clock, and suppose that,
instead of sending~$\mu$,
Alice sends at time~$m$
a message~$\mu'$ that specifies the sending time,
such as
$$\mbox{
``This message is being sent at time~$m$; $\mu$.''
}$$
Recall that it is common knowledge that every message sent by Alice
is received by Bob within~$\eps$ time units.
When Bob receives~$\mu'$, he knows that~$\mu'$ was sent at
time~$m$.  Moreover,
Bob's receipt of~$\mu'$
is guaranteed to happen no later than time $m+\eps$.
Since Alice and Bob use the same clock, it is common knowledge at
time $m+\eps$ that it is $m+ \eps$.
It is also
common knowledge that any message sent at time~$m$ is received by time
$m+\eps$.  Thus, at
time~$m+\eps$,
the fact that Alice sent~$\mu'$ to Bob
is common knowledge.

Note that in the first example
common knowledge will never hold regardless of whether~$\eps$ is a
day, an hour, or a nanosecond. The slight uncertainty about the
sending time and the message transmission time prevents common
knowledge of~$\mu$ from ever being attained in this scenario.
What makes the second example so
dramatically different?
When a fact~$\phi$ is common knowledge,
everybody must know that it is.
It is impossible for agent~$i$ to know that~$\phi$ is common knowledge
without agent~$j$ knowing it as well.
This means that the transition
from~$\phi$ not being common knowledge to its being common
knowledge must involve a {\em simultaneous} change in all relevant
agents' knowledge.  In the first example, the uncertainty makes  such a
simultaneous transition impossible, while in the second,
having the same clock makes a simultaneous transition
possible and this transition occurs at time $m+\eps$.
These two examples help
illustrate the connection between simultaneity and common knowledge
and the effect this can have on the attainability of common knowledge.
We now formalize and further explore this connection.%
\index{common knowledge!and uncertainty|)}

\section{Simultaneous Events}\label{simultaneous}
The Alice and Bob examples
illustrate how
the transition from a situation in which a fact is not common
knowledge to one where it is common knowledge
requires simultaneous events to take place at all sites of the system.
The relationship between simultaneity and common knowledge,
is in fact even more fundamental than that.
We saw by example earlier that actions that must
be performed simultaneously by all parties,
such as attacking in the coordinated attack problem,%
\index{coordinated attack}
become common knowledge as soon as they are performed:
common knowledge is a prerequisite for
simultaneous actions.
In this section, we give a result that says
that a fact's becoming common knowledge
requires the occurrence of simultaneous events at different sites of
the system.
Moreover, the results say that
in a certain technical sense,
the occurrence of simultaneous events is necessarily common
knowledge.
This demonstrates
the strong link between common
knowledge and simultaneous events.

To make this claim precise,
we need to formalize the notion
of simultaneous events.
We begin by
briefly reviewing the framework of \cite{FHMV} for modeling multi-agent
systems.%
\footnote{The general framework presented here for ascribing knowledge
in multi-agent systems originated with Halpern and Moses
\cite{HM1,Mosthesis} and
Rosenschein \cite{Ros}.
Variants were also introduced by
Fischer and Immerman \cite{FI1},
Halpern and Fagin \cite{HF87},
Parikh and Ramanujam
\cite{PR}, and Rosenschein and Kaelbling \cite{RK}.}
We assume that at each point in time, each agent is in some {\em
local state}.
Informally, this local state encodes the information available
to the agent at this point.
In addition, there is an {\em environment\/} state, that
keeps track of everything relevant to the system not recorded in the
agents' states.

A {\em global state\/} is an $(n+1)$-tuple
$(s_e,
s_1, \ldots, s_n)$ consisting of
the environment state $s_e$ and
the local state $s_i$
of each agent $i$.
A {\em run\/} of the system is a function from time
(which, for
ease of exposition, we assume ranges over the natural numbers)
to global states.
Thus, if $r$ is a run, then $r(0), r(1), \ldots$ is a sequence of
global states that, roughly speaking, is a complete description of
how the system evolves over time
in one possible execution of the system.
We take a {\em system\/}
to consist of a set of runs.  Intuitively, these runs describe
all the possible sequences of events that could occur in a system.

Given a system $\R$,
we refer to a pair $(r,m)$ consisting of a run $r \in \R$ and a time
$m$ as a {\em point}.
If $r(m) =
(s_e,
s_1, \ldots, s_n)$, we define
$r_i(m) = s_i$\glossary{$r_i(m)$},
for
$i = 1, \ldots, n$; thus, $r_i(m)$ is process~$i$'s
local state at the point $(r,m)$.
We say two points $(r,m)$ and
$(r',m')$
are {\em indistinguishable\/} to agent $i$, and write $(r,m) \sim_i
(r',m')$, if
$r_i(m) = r'_i(m')$,
\ie if agent $i$ has the same local state at both points.
Finally, we define an
{\em interpreted system\/} to be a pair $(\R,\pi)$ consisting of a
system $\R$ together with a mapping $\pi$ that associates a truth
assignment to the primitive propositions with each global state.

An interpreted system can be viewed as a Kripke structure: the
points are the possible worlds, and $\sim_i$ plays the role of
the accessibility relation.
We give semantics to knowledge formulas in interpreted systems just
as in Kripke structures:
Given a point $(r,m)$ in an interpreted system $\I = (\R,\pi)$, we have
$(\I,r,m) \sat K_i \phi$ if $(\I,r',m') \sat \phi$ for all points
$(r',m')$ such that $(r',m') \sim_i (r,m)$.
Notice that under this
interpretation,
an agent knows $\phi$ if $\phi$ is true at
all the situations the system could be in, given the agent's
current information
(as encoded by his local state).
Since
$\sim_i$ is an equivalence relation, knowledge
in this framework satisfies the
axioms of the modal system S5.
If $G$ is a set of agents, we define
$E_G$\glossary{\gloseg}
(``everyone in the group $G$ knows'')
by saying
$(\I,r,m) \sat \EG \phi$ if $(\I,r,m) \sat K_i \phi$
for every $i \in G$.
We define
$C_G$\glossary{\gloscg}
(``it is common knowledge among the agents in~$G$'') by saying
$(\I,r,m) \sat \CG \phi$ if $(\I,r,m) \sat
(E_G)^k \phi$ for every $k$.
When $G$ is the set of all agents, we may write $E$ for $E_G$,
and $C$ for $C_G$.
We write $\I \sat \phi$ if $(\I,r,m) \sat \phi$ for every
point $(r,m)$ of the system $\I$.

We now give a few more
definitions, all relative to a fixed interpreted
system~$\cI=(\R,\pi)$.%
\index{interpreted system}
Let~$S$ denote the set of points of the system~$\R$.
Define an {\em event\/}\index{event} in~$\R$
to be a subset of~$S$;
intuitively, these are the points where the event~$e$ holds.
An event~$e$ is said to
{\em hold\/}\index{event!holds}
at a point~$(r,m)$ if $(r,m) \in e$.
Of special interest are events whose occurrence is reflected in an
agent's local state.
More formally, an event~$e$ is {\em local to~$i$}%
\index{event!local}
(in interpreted system~$\I$)
if there is a set~$L^e_i$
of~$i$'s local states such that for all points~$(r,m)$
we have $(r,m) \in e$ iff $r_i(m)\in L^e_i$.
The events
of sending a message,
receiving a message, and performing an internal action
are examples of local events for agent~$i$.
We remark that the definition of a local event does not imply that an
event that is local to~$i$ cannot also be local to~$j$. In order to be
local to both agents, it only needs to be reflected in the local
states of
both
agents.

Certain events depend only on the global state.
An event~$e$ is a {\em state event\/}%
\index{state event}
if there is a set $\G^e$ of global states such that
for all points~$(r,m)$
we have $(r,m) \in e$ iff $r(m)\in \G^e$.
It is easy to see that local events are state events.
More generally,
a state event is one that depends only on what is recorded
in the local states of the agents and the state of the environment.
We associate with every state event~$e$ a primitive
proposition~$\psi_e$\glossary{\glospsie}
that is true at the global state $r(m)$ if and only if
$(r,m) \in e$.
This is well-defined, because
it follows easily from the definition
of state events that if
$e$ is a state event and
$(r,m)$ and $(r',m')$ are points
such that
$r(m) = r'(m')$,
then $(r,m) \in e$ if and only if $(r',m') \in e$.

We can similarly associate with every formula~$\phi$ an
event
$\evi(\phi) = \{(r,m) : (\cI,r,m)\sat\phi\}$.
The event $\evi(\phi)$ thus holds exactly when~$\phi$ holds.
We call~$\evi(\phi)$ {\em the event of~$\phi$
holding (in~$\I$)}.
It is easy to check that an event~$e$ is local to~$i$
if and only if $K_i \psi_e$ holds, that is,
if and only if $i$ knows
that~$e$ is holding.
Moreover, the event of $K_i\phi$ holding is always a local event
for~$i$.

We are now ready to address the issue of simultaneous events.
Intuitively, two events are simultaneous if they occur at the
same points.
Our interest in simultaneity
is primarily in the context
of coordination.
\index{coordination!perfect|(}
Namely, we are interested in events that are local to different
agents and are coordinated in time.
Thus, we concentrate on events
whose occurrence is simultaneously reflected in the local state of
the agents.
More formally, we define
an {\em event ensemble for~$G$}%
\oldindex{event!ensemble}
(or just {\em ensemble}\index{ensemble} for short)
 to be a mapping $\e$
assigning to every agent~$i\in G$ an event~$\e(i)$ local to~$i$.
An ensemble~$\e$ for~$G$ is said to be
{\em perfectly coordinated}%
\index{ensemble!perfectly coordinated|(}
if the local events in $\e$ hold simultaneously; formally,
if $(r,m) \in \e(i)$ for some $i \in G$, then $(r,m) \in \e(j)$
for all $j \in G$.
Thus, the ensemble $\e$ for~$G$ is perfectly coordinated
precisely if $\e(i)=\e(j)$ for all $i,j\in G$.
Since an event $e$ that is local to agent~$i$
is defined in terms of a set $L^{e}_i$ of states
local to agent~$i$, the ensemble $\e$ for~$G$
is perfectly coordinated if all the agents in~$G$ enter
their respective sets~$L^{\e(i)}_i$ simultaneously.
Thus, the events in a perfectly coordinated ensemble are
simultaneous.

An example of a perfectly coordinated ensemble
is the set of local events that correspond to
the ticking of a global clock,%
\index{clock}
if the ticking
is guaranteed to be reflected simultaneously at all sites of a system.
Another example is the event of shaking hands: being a mutual action,
the handshakes of the parties are perfectly coordinated.

Given an ensemble~$\e$ for~$G$, the proposition
$\psi_{\e(i)}$ corresponds to the state event $\e(i)$ holding.
We
also
define $\psi_{\e}=\bigvee_{i\in G}\psi_{\e(i)}$.
Thus, $\psi_{\e}$
is true whenever one of the state events $\e(i)$ holds.

\begin{prop}
\label{p--simck}
Let $\cI$ be  an interpreted system and~$G$ a set of agents.
\begin{enumerate}
\item[(a)] For every formula~$\phi$,
the ensemble~$\e$ for~$G$ defined by
$\e(i)=\evi({K_i\CG\phi})$
is perfectly coordinated.
\item[(b)] If~$\e$ is a perfectly coordinated ensemble for~$G$,
then $\cI\sat \psi_{\e}\imp\CG\psi_{\e}$.
\end{enumerate}
\end{prop}

\noindent
(In fact, $K_i C_G \phi$ in part
(a) of Proposition~\ref{p--simck} is
logically
equivalent to $\CG \phi$, but we write $K_i \CG \phi$
to bring out the similarities between this result and
Proposition~\ref{p--eps-ck} below.)
Proposition~\ref{p--simck} precisely captures the close correspondence
between common knowledge and simultaneous events.
It asserts that the local events that correspond to common knowledge
are perfectly coordinated, and the local events in a perfectly
coordinated ensemble are common knowledge when they hold.
Notice that part~(a) implies in particular that the
transitions from
$\neg K_i \CG\phi$ to~$K_i\CG\phi$,
for $i \member G$, must be simultaneous.
Among other things, this
helps clarify the difference between the two examples considered
in Section~\ref{ck-and-unc}:
In the first example,
Alice and Bob cannot attain common knowledge of~$\snt(\mu)$
because they are unable to make such a simultaneous transition,
while in the second example they can (and do).

The close relationship between common knowledge and simultaneous actions
is what makes common knowledge such a useful tool for analyzing
tasks involving coordination and agreement.
It also gives us some insight into how common knowledge arises.
For example, the
fact that a public announcement has been made is common knowledge,
since the announcement is heard simultaneously by everyone.
(Strictly speaking, of course, this is
not quite true; we return to this issue in Section~\ref{simult-model}.)
More generally, simultaneity is inherent in the notion of
{\em copresence}.%
\index{copresence}
As a consequence,
when people sit around a table,
the existence of the table, as well as the nature of the objects on the
table, are common knowledge.

Proposition~\ref{p--simck} formally captures the role of
simultaneous actions in making agreements%
\index{agreement}
and conventions common knowledge.
As we discussed earlier,
common knowledge is inherent in agreements and conventions.
Hand shaking, face-to-face or telephone conversation,
and a simultaneous signing of a contract are
standard ways of reaching agreements. They all involve simultaneous
actions and have the effect of making the agreement common knowledge.

\section{Temporal Imprecision}\label{imprecision}

As we illustrated previously
and formalized in Proposition~\ref{p--simck},
simultaneity is inherent in the notion of common
knowledge (and vice versa).
It follows that simultaneity is a prerequisite for
attaining common knowledge.
Alice and Bob's failure to reach
common knowledge in the first example above can therefore be blamed
on their inability to perform a simultaneous state transition.
As might
be expected, the fact that simultaneity is a prerequisite for
attaining common knowledge has additional consequences.
For example, in many distributed systems%
\index{distributed system}
each process possesses a clock.%
\index{clock}
In practice, in any distributed system there is always some
uncertainty regarding the relative synchrony of the clocks
and regarding the precise message transmission times.
This results in what is called the
{\em temporal imprecision}%
\index{temporal imprecision|(} of the
system. The amount of temporal imprecision in different systems varies,
but it can be argued that every practical system will have some
(possibly very small) degree of imprecision. Formally, a given
system~$\R$ is said to have
{\em temporal imprecision}\index{temporal imprecision}
if for all runs~$r\in \R$, times~$m$, and sets~$G$ of processes with
$|G|\ge 2$, there exist processes $i,j\in G$ with $i\ne j$, a
run $r' \in \R$, and a time $m'$ such that
$r'_i (m')=r_i(m)$ while $r'_j (m')=r_j(m+1)$.
Intuitively, in a system with temporal imprecision, $i$~is
uncertain about $j$'s clock reading\index{clock}; at the point $(r,m)$,
process~$i$ cannot tell whether $j$'s clock is characterized by $j$'s
local state at $(r,m)$ or $j$'s local state at $(r,m+1)$.
Techniques from the distributed-systems%
\index{distributed system}
literature \cite{DHS,HMM}
can be used to show that any system in
which, roughly speaking, there is some
initial uncertainty regarding relative clock readings and uncertainty
regarding exact message
transmission times must have temporal imprecision.

Systems
with temporal imprecision turn out to have the property that
no protocol can
be guaranteed
to synchronize the processes'
clocks perfectly.%
\index{clock}
As we now show, events cannot be perfectly coordinated in systems
with temporal imprecision
either.  These two facts are closely related.

We define an
ensemble~$\e$
for~$G$
in~$\I$ to be
{\em nontrivial}\index{ensemble!nontrivial} if
there exist a run~$r$ in~$\I$ and times~$m,m'$ such that
$(r,m)\in \union_{i\in G} \e(i)$ while $(r,m')\notin \union_{i
\in G} \e(i)$.
Thus, if $\e$ is a perfectly coordinated ensemble
for~$G$, it is
{\em trivial\/}\index{ensemble!trivial}
if for each run~$r$ of the system and for each agent $i \in G$,
the events in~$\e(i)$ hold either at all points of~$r$ or at no
point of~$r$.
The definition of systems with temporal imprecision implies
the following:

\begin{prop}
\label{p--no-se}
In a system with temporal imprecision
there are no nontrivial perfectly coordinated ensembles
for~$G$, if $|G| \ge 2$.
\end{prop}
We thus have the following corollary.
\begin{cor}
\label{c--nock}
{\rm \cite{HM1}}
Let~$\cI$ be a system with temporal imprecision, let~$\phi$
be a formula, and let $|G|\ge 2$. Then for all runs~$r$ and times~$m$
we have $(\cI,r,m)\sat\CG\phi$ iff $(\cI,r,0)\sat\CG\phi$.
\end{cor}

In simple terms, Corollary~\ref{c--nock} states that no fact can
become common knowledge during a run of a system with temporal
imprecision.
If the units by which time%
\index{time}
is measured in our model are sufficiently
small, then all practical distributed systems%
\index{distributed system}
have temporal imprecision.
For example, if we work at the nanosecond level, then there is bound to
be some uncertainty regarding exact message transmission times.  On the
other hand, if we model time at the level of minutes, this uncertainty
may disappear.
As~a result, Corollary~\ref{c--nock} implies that
no fact can ever become common knowledge in practical
distributed systems.
Carrying this argument even further, we can view essentially all
real-world scenarios as
ones
in which true simultaneity cannot be guaranteed.
For example, the children in the
muddy children puzzle\index{muddy children puzzle}
neither hear nor comprehend the father simultaneously.
There is bound to be some uncertainty about how long it takes
each of them to process the information.
Thus, according to our earlier discussion, the children in fact
do not attain common knowledge of the father's statement.

We now seem to have a paradox.
On the one hand, we have argued that common knowledge is unattainable in
practical contexts.  On the other hand, given our claim that
common knowledge is a prerequisite for agreements and conventions
and the observation that we do reach agreements and
that conventions
are maintained, it seems that common knowledge {\em is\/}
attained in practice.

What
is the catch?  How can we explain this discrepancy between
our practical experience and our technical results?
In the next two sections,
we consider two resolutions to
this paradox.  The first rests on the observation that if we model time
at a sufficiently coarse level, we can and do attain common knowledge.
The question then becomes when and whether it is
appropriate to model time in this way.  The second
says that, although we indeed cannot attain common knowledge,
we can attain close approximations of it,
and this suffices for our purposes.

\section{The Granularity of Time}
\label{simult-model}
Given the complexity of the real world, any
mathematical model of a situation must abstract away
many details.  A useful model is typically one that
abstracts away  as much of the irrelevant detail as possible, leaving
all and only the relevant aspects of a situation.
When modeling a particular situation,
it can often be quite difficult
to decide the level of granularity at which to model time.
The notion
of time\index{time} in a run rarely corresponds to real time.
Rather, our choice of the granularity of time is motivated by
convenience of modeling.  Thus, in a distributed application,%
\index{distributed system}
it may be perfectly appropriate to take a round to be sufficiently long
for a process to send a message to all other processes, and perhaps
do some local computation as well.

As we have observed,
the argument that every practical system has
some degree of temporal imprecision holds
only relative
to a
sufficiently fine-grained model of time.
For Proposition~\ref{p--no-se}
and Corollary~\ref{c--nock} to apply, time must be represented in
sufficiently fine detail for temporal imprecision
to be reflected in the model.
If a model has a coarse notion of time, then
simultaneity, and hence common knowledge, are often attainable.
For example, in synchronous systems
(those where the agents have access to a shared clock, so that,
intuitively, the time is common knowledge)
there is no temporal imprecision.%
\index{temporal imprecision}
As an example,
consider a simplified model of the muddy children problem.
The initial states of the
children and the father
describe what they see; later states describe
everything they have heard.
All communication proceeds in rounds.
In round~1, if there is at least one muddy child, a
message to this effect is sent to all children.
In the
odd-numbered rounds~1, 3, 5, \ldots, the father
sends to all children the message
``Does any of you know whether you have mud on your own forehead?''
The children respond ``Yes'' or ``No'' in the
even-numbered rounds.
In this simplified model,
the children do attain common knowledge of the father's statement
(after the first round).
If, however, we ``enhance'' the model to take into
consideration the minute details of the neural activity in the
children's brains, and considered time on, say, a millisecond scale,
the children would not be modeled as hearing the father
simultaneously. Moreover, the children would not attain common
knowledge of the father's statement.
We conclude that whether a given fact
becomes common knowledge at a certain point, or in fact
whether it {\em ever} becomes common knowledge, depends in a crucial way
on the model being used.
While common knowledge may be attainable in
a certain model of a given
real world situation, it becomes unattainable once we consider
a more detailed model
of {\em the same situation}.

When are we
justified in reasoning
and acting as if common knowledge is attainable?
This reduces to the question of when we can argue that one
model---in our case a coarser or
less detailed model---is ``as good'' as another, finer,  model.
The answer, of course, is ``it depends on the intended application.''
Our approach
for deciding whether a
less detailed model is as good as another, finer,  model,
is to assume that there is some ``specification'' of interest, and to
consider whether the finer model satisfies the same
specification as the coarser model.
For example, in the muddy children puzzle, our earlier model
implicitly assumed that
the children all hear the father's initial
statement
and his later questions simultaneously.
We can think of this as a coarse
model where, indeed, the children attain common knowledge.
For the fine model,
suppose instead
that every time the father
speaks, it takes somewhere between 8 and 10 milliseconds for each
child to hear and process what the father says,
but the exact time may be different for each child, and may even
be different for a given child every time the father speaks.
Similarly, after a given child speaks, it takes between 8 and 10
milliseconds for the other children and the father to hear and
process what he says.
(While there is nothing particularly significant in our choice of
8 and 10 milliseconds, it is important that a child does not hear
any other child's response to the father's question before he utters
his own response.)
The father does not ask his $k^{\rm th}$ question until he has received
the responses from all children
to his $(k-1)^{\rm st}$ question.

The specification of interest for the muddy children puzzle is
the following:
A child says ``Yes'' if he knows
whether he is muddy and says ``No'' otherwise.
This specification is satisfied in particular when each child
follows the protocol that if he sees
$k$ muddy children, then he
responds ``No'' to the father's first~$k$
questions and ``Yes'' to
all the questions after that.
This specification is true in both the coarse model and the fine
model.
Therefore, we consider the coarse model adequate.
If part of the specification had been that the children answer
simultaneously, then the coarse model would not have been adequate.
For a more formal presentation of
our
approach, see \cite{FHMV}.

The observation that whether or not common knowledge is attainable
depends in part on how we model time was made
in a number of earlier papers \cite{Aumann89,FI1,HM1,Kurki,Nei,NT}.
Our approach formalizes this observation and offers a rigorous
way to determine when the coarse model is
adequate.

\section{Approximations of Common Knowledge}\label{s--variants}

Section~\ref{simultaneous} shows that common knowledge captures the
state of knowledge resulting from simultaneous events.
It also shows, however, that in the absence of events that
are guaranteed to hold simultaneously, common
knowledge is not attained.
In Section~\ref{simult-model},
we tried to answer the question of when we can
reason and act as if certain events were simultaneous.
But there is another point of view we can take.
There are situations where events holding at different
sites need not happen simultaneously;
the level of coordination required is
weaker than absolute simultaneity.
For example, we may want the events to hold at most a certain
amount of time apart.
It turns out that just as common knowledge is the state of knowledge
corresponding to perfect coordination, there are states of shared
knowledge corresponding to other forms of coordination.
We can view these states of knowledge as approximations of true
common knowledge.
It is well known that common knowledge can be defined in
terms of a fixed point, as well as an infinite conjunction.
As shown in \cite{HM1}, $C_G\phi$ is equivalent to
$\nu x [E_G(\phi \land x)]$,
where $\nu x$ is the {\em greatest fixed-point operator}.%
\footnote{Formal definitions of this operator can be found in
\cite{FHMV,HM1}.}
As we shall see, the
approximations
of common knowledge have similar
fixed-point definitions.  Fortunately, while perfect coordination
is hard to
attain in practice, weaker forms of coordination are often attainable.
This is one explanation as to
why the unattainability of common knowledge might not spell as
great a disaster as we might have originally expected.
This section considers two of these weaker forms of coordination,
and their corresponding states of knowledge.
\index{simultaneity|)}

Let us return to the first Alice and Bob example.
Notice that if~$\eps=0$, then Alice and
Bob attain common knowledge of~$\snt(\mu)$ immediately after the message
is sent.  In this case, it is guaranteed that once the message is
sent, both agents immediately know the contents of the message, as
well as the fact that it has been sent.
Intuitively, it seems that the closer~$\eps$ is to~0,  the closer
Alice and Bob's state of knowledge should be to common knowledge.
Compare the situation when~$\eps>0$ with~\mbox{$\eps=0$}.
As we saw, if~$\eps>0$ then Alice does
not know that Bob received her message immediately after she sends the
message.
She does, however, know that {\em within $\eps$ time units} Bob will
receive the message and know both the contents of the message and
that the message has been sent.
The sending of the message results in a situation where, within $\eps$
time units,
everyone knows that the situation holds.
This is analogous to the fact that
common knowledge corresponds to a situation
where
everyone knows that the situation holds.
This suggests that the state of knowledge resulting in the Alice and
Bob scenario should involve a fixed point of some sort.
We now formalize a notion of coordination related to the Alice and Bob
example, and define an approximation of common knowledge corresponding
to this type of coordination.

\index{coordination!epsilon-@$\eps$-|(}
An ensemble~$\e$ for~$G$
is said to be {\em $\eps$-coordinated}%
\index{ensemble!epsilon-coordinated@$\eps$-coordinated|(}
(in a given system~$\cI$) if the local events in $\e$
never hold more than~$\eps$ time units apart;
formally, if $(r,m) \in \e(i)$ for some $i \in G$,
then there exists an interval~$I=[m',m'+\eps]$ such that
$m\in I$ and for all $j\in G$ there exists~$m_j\in I$ for which
$(r,m_j)\in \e(j)$.
Note that $\eps$-coordination with $\eps=0$ is perfect coordination.
While it is essentially infeasible in practice to coordinate
events so that they hold simultaneously at different sites of a
distributed system,
$\eps$-coordination is often attainable in practice,
even in systems where there is uncertainty in message delivery time.
Moreover, when~$\eps$ is sufficiently small, there are many
applications for which $\eps$-coordination is practically as good as
perfect coordination.
For example, instead of requiring
a simultaneous attack in the coordinated attack problem,
it may be sufficient to
require only that the two divisions attack
within a certain $\eps$-time bound of each other.
This is called an $\eps$-{\em coordinated attack}.

More generally,
$\eps$-coordination may be practically as good as
perfect coordination for many instances of
agreements and conventions.
One example of $\eps$-coordination results from a message being
broadcast to all members of a group~$G$, with the guarantee that it
will reach all of the members within~$\eps$ time units of one another.
In this case it is easy to see that when an agent receives the message,
she knows the message has been broadcast, and knows that
within~$\eps$ time units each of the members of~$G$ will have
%
 received the message and will know that within~$\eps$ \dots

Let $\eps$ be arbitrary.
We say that {\em within an
$\eps$~interval everyone in~$G$ knows~$\phi$\/}, denoted
$\Ee\subG\phi$,\glossary{\glosege}
if there is an interval of~$\eps$ time units
containing the current time such that each process comes to know~$\phi$
at some point in this interval. Formally,
$(\I,r,m)\sat\Ee\subG\phi$
if there exists an interval~$I=[m',m'+\eps]$ such that
$m\in I$ and for all $i\in G$ there exists~$m_i\in I$ for which
$(\I,r,m_i)\sat K_i\phi$.
Thus, in the case of Alice and Bob, we have
$\I\sat\snt(\mu)\imp\Ee_{\{A,B\}}\snt(\mu)$.
We
define $\eps$-common knowledge,
denoted by~$\Ce\subG$, using a greatest fixed-point operator:
$\Ce_G\phi \eqdef \nu x[\Ee_G(\phi\wedge x)]$.
Notice how similar this definition is to
the
fixed-point definition of common knowledge.
The only change is in
 replacing~$\EG$ by~$\Ee\subG$.

Just as common knowledge is closely related to
perfect coordination, $\eps$-common knowledge is
closely related to
$\eps$-coordination.
We now make this claim precise.
The next proposition is
analogous to
Proposition~\ref{p--simck}.

\begin{prop}
\label{p--eps-ck}
Let $\cI$ be  an interpreted system and~$G$ a set of agents.
\begin{enumerate}
\item[(a)]
For every formula~$\phi$,
the ensemble~$\e$ for~$G$ defined by
$\e(i)=\evi({K_i\Ce_G\phi})$
is $\eps$-coordinated.
\item[(b)]
If~$\e$ is an $\eps$-coordinated ensemble for~$G$,
then $\cI\sat \psi_{\e}\imp \Ce_G\psi_{\e}$.
\end{enumerate}
\end{prop}
Note that in part~(a), we write $K_i \Ce_G\phi$; we cannot write
$\Ce_G\phi$, since $\evi(\Ce_G\phi)$ is not an event local to agent $i$.

Since in the coordinated attack problem message delivery is not
guaranteed, it can be shown that the generals cannot
achieve even
$\eps$-coordinated attack.
On the other hand,
if messages are
guaranteed to be delivered within $\eps$ units of time, then
$\eps$-coordinated attack can be accomplished.
General~$A$ simply sends General~$B$ a message saying ``{\sl attack}''
and attacks immediately; General~$B$ attacks upon receipt of
the message.

Although $\eps$-common knowledge is useful for the analysis of systems
where the uncertainty in message communication time is small, it is
not quite as useful in the analysis of
systems where message delivery
may be delayed for a long period of time.
In such systems,
rather than perfect or $\eps$-coordination, what can often be
achieved is {\em eventual\/} coordination.
An ensemble~$\e$ for~$G$ is {\em eventually coordinated\/}
(in a given system~$\cI$) if,
for every run of the system, if some event in~$\e$ holds during the
run, then all events in~$\e$ do.
More formally, if $(r,m) \in \e(i)$ for some $i \in G$, then
for all $j\in G$ there exists some $m_j$ for which
$(r,m_j)\in \e(j)$.
An example of an eventual coordination of~$G$ consists of the
delivery of (copies of) a message broadcast to every member of~$G$
in a system with message delays.
An agent receiving this message knows the
contents of the message, as well as the fact that each other member
of~$G$ must receive the message at some point in time, either past,
present, or future.

Eventual coordination
gives rise to {\em eventual\/}%
\index{eventual common knowledge|(}
common knowledge, denoted by $\Cd_G$,\glossary{\gloscgd}
and defined by
$\Cd_G\phi\eqdef \nu x[\Ed_G(\phi\wedge x)]$.
Here we define $\Ed_G\phi$\glossary{\glosegd}
to hold at $(\I,r,m)$ if for
each $i\in G$ there is some time~$m_i$ such that
$(\I,r,m_i)\sat K_i\phi$.
Thus, $\Ed_G$ can be viewed as the limit of $\Ee_G$ as $\eps$
approaches infinity.
It is straightforward to show that~$\Cd_G$ is related to eventual
coordination just as~$\CG$ is related to
perfect coordination,
and~$\Ce_G$ to $\eps$-coordination.
Interestingly, although $\Ce_G$ is definable as an infinite conjunction,
it can be shown that $\Cd_G$ is not \cite{FHMV}.
We really need to use fixed points here;
cf.~\cite{Bar1}.

Just as $\eps$-coordinated attack is a weakening of the simultaneity
requirement of coordinated attack, a further weakening of the
simultaneity requirement is given by
{\em eventually coordinated attack}.
This requirement says that if one of the two divisions attacks, then
the other division eventually attacks.
If messages are guaranteed to be delivered eventually,
then even if there is no bound on message delivery time,
an eventually coordinated attack can be carried out.

The notions of $\eps$-common knowledge and of eventual
common knowledge are from \cite{HM1}.
Our contribution here is in introducing ensembles as a formalization
of the concept of coordination and in showing that approximations of
common knowledge correspond to approximations of coordination.
We note also that other approximations to common knowledge have
been considered, including timestamped common knowledge \cite{HM1},
probabilistic common knowledge \cite{BD87,FH3,HT,KPN90,MSamet},
and concurrent common knowledge \cite{PT}.
All these can be defined via small variations on the fixed-point
definition of common knowledge.
All of these variants are weaker than common knowledge.
The state of continual common knowledge defined and used
in \cite{HMW} is a variant of common knowledge that is generally
strictly {\em stronger} than common knowledge.

\section{Summary}

The central theme of this paper is an attempt to resolve
the paradox of common knowledge raised in \cite{HM1}:
Although
common knowledge
can be shown to be a prerequisite for
day-to-day activities of coordination and agreement,
it can also be shown to be unattainable in practice.
The resolution of this paradox leads to
a deeper understanding of the nature of common knowledge and
simultaneity, and shows once again the importance of the modeling
process.
In particular, it brings out the importance of the granularity at which
we model time, and stresses
the need to consider the
applications for which these notions are being used.
Moreover, by using the notion of event ensembles, we are able to
clarify the tight relationship between common knowledge and
coordination.

\end{document}